# Anisotropic MHD Turbulence

C. S. Ng and A. Bhattacharjee

*Space Science Center, Institute for the Study of Earth, Oceans, and Space, University of New Hampshire, Durham, NH 03824*

**Abstract.** The solar wind and the interstellar medium are permeated by large-scale magnetic fields that render magnetohydrodynamic (MHD) turbulence anisotropic. In the weak-turbulence limit in which three-wave interactions dominate, analytical and high-resolution numerical results based on random scattering of shear-Alfvén waves propagating parallel to a large-scale magnetic field, as well as direct simulations demonstrate rigorously an anisotropic energy spectrum that scales as $k_\perp^{-2}$, instead of the famous Iroshnikov-Kraichnan (IK) spectrum of $k^{-3/2}$ for the isotropic case. Even in the absence of a background magnetic field, anisotropy is found to develop with respect to the local magnetic field, although the energy spectrum is globally isotropic and is found to be consistent with a $k^{-3/2}$ scaling. It is also found in direct numerical simulations that the energy cascade rate is much closer to IK scaling than a Kolmogorov scaling. Recent observations in the solar wind on cascade rates (as functions of the proton temperature and solar wind speed at 1 AU) seem to support this result [Vasquez et al. 2007].

**Keywords:** Magnetohydrodynamic turbulence, solar wind, interstellar medium, energy cascade

**PACS:** 52.35.Ra, 52.35.Mw

## INTRODUCTION

Incompressible magnetohydrodynamics (MHD) has been the standard point of departure for theoretical studies of interstellar medium (ISM) and solar wind turbulence. Although the applicability of the model to the high-beta and weakly collisional plasmas commonly encountered in the ISM and solar wind is open to serious questions, the predictions of this model must be understood before one can attempt a more complete dynamical theory that includes compressibility and kinetic effects. Since MHD equations reduce to the incompressible Navier-Stokes equation in the limit of zero magnetic field, it is natural to expect that a theory of MHD turbulence will be an extension of the highly successful Kolmogorov theory in hydrodynamics (HD), but with the effects of magnetic fields included [1]. The Kolmogorov theory uses simple dimensional analysis and predicts the well-known one-dimensional energy spectrum $E(k) \propto k^{-5/3}$ where $k$ is the wavenumber. The theory relies on two fundamental assumptions: turbulence is isotropic and the energy cascade is dominated by interaction between eddies of similar spatial scales. The Iroshnikov-Kraichnan (IK) theory [2,3] of incompressible MHD turbulence attempts to build on the successful framework of Kolmogorov, including the effect of a magnetic field. In the IK theory, small-scale fluctuations are envisioned to behave as Alfvén wave packets propagating along local magnetic field lines (with or without a uniform background magnetic field). The collision between two oppositely propagating wave packets then provides a mechanism for the cascade of energy to short wavelengths. This picture of colliding wave packets seems to be supported by evidence from numerical experiments, as well as *in situ* observations of the solar wind. While the IK theory elucidates the role of colliding Alfvén wave packets in the generation of the cascade, it retains the assumptions of isotropy and local interactions in $k$-space implicit in the Kolmogorov theory.

Unlike the energy spectrum in the Kolmogorov theory, the energy spectrum in the IK theory cannot be determined entirely by dimensional analysis because it contains an additional dependence on the macroscopic magnetic field. To determine this dependency, IK argued that three-wave interactions between colliding Alfvén wave packets dominate the energy cascade in the inertial range, assumed to be isotropic. Under these assumptions, the energy spectral index is found to be 3/2, in both 2D and 3D turbulence.

The violation of the assumption of isotropy in the presence of background magnetic field has been recognized by several investigators (see ref. [4] and references there in). In the strong turbulence regime, it has been proposed that anisotropy develops until a condition of "critical balance" is attained whereby the Alfvén time becomes of the same order as the eddy turnover time [5]. Under these conditions, scaling arguments show that the anisotropy becomes scale-dependent and one recovers an anisotropic Kolmogorov spectrum for MHD turbulence. However, it is not clear if MHD turbulence can be described as strong in all situations. If the weak turbulence assumption is valid, it has been shown by scaling argument [4,6], a random wave collision model [7], as well as kinetic wave equation [8] that the energy spectrum has $k_\perp^{-2}$ dependence in the perpendicular direction. We will present more support of this in the next section using results from direct numerical simulations.

The considerations described in the previous paragraph for anisotropic MHD turbulence applies when a uniform background magnetic field is assumed to be present. This is *not* the case considered in the original IK theory. However, local anisotropy relative to the local magnetic field can still exist, which raises serious questions about the validity of the IK theory [9,10]. However, there appears to be no consensus yet on the precise scaling of the anisotropic energy spectrum. Our recent 2D MHD simulations seem to give a spectral index of 3/2 rather than 5/3. However, since the difference between these two numbers is small, this may not be considered conclusive. Another way to look at the difference between the Kolmogorov and IK scalings is by comparing their energy cascade rates with numerically obtained rate. We will present such result later in this paper.

## WEAK TURBULENCE WITH UNIFORM MAGNETIC FIELD

The original IK theory was developed for the case without a uniform background magnetic field $\mathbf{B}_0$. With that, the energy spectrum $E(k_\parallel, k_\perp)$ becomes globally anisotropic. (Here $k_\parallel$ and $k_\perp$ are the parallel and perpendicular component of the wave vector $\mathbf{k}$ with respect to $\mathbf{B}_0$.) Dimensional analysis cannot provide a definite scaling because it cannot discriminate between the two length-scales perpendicular ($k_\perp^{-1}$) and parallel ($k_\parallel^{-1}$), for the full 3D spectrum. We must then estimate the spectral index by invoking additional physical arguments.

In ideal incompressible MHD, a wave packet moving in one direction along $\mathbf{B}_0$ is an exact solution and thus energy cascade can happen only when it collides with another wave packet moving in the opposite direction. Now, the collision time-scale between wave packets of scale $k_\parallel$ is $\tau_k \sim 1/k_\parallel V_A$. The change in amplitude in one collision is small and thus it takes many collisions (of the order of $\chi^{-2}$, where $\chi \equiv k_\perp v_{k\perp}/k_\parallel V_A$ is a small parameter) to cascade energy. As three-wave interactions dominate for weak turbulence, the IK analysis can thus be repeated, which yields $\varepsilon \sim k_\perp^4 k_\parallel E^2(k_\perp)/V_A$ in a direction perpendicular to $\mathbf{B}_0$ and $k_\parallel$ treated as a parameter. Thus the anisotropic spectrum becomes $E(k_\perp) \propto \varepsilon^{1/2} k_\perp^{-2}$ [4,6]. Since $\varepsilon$ is inversely proportional to $V_A$, the large-scale field has the effect of reducing energy cascade.

Using the 3D reduced MHD (RMHD) equation and based on the analytic expression of three-wave interactions [11], one can calculate the change in a wave packet for multiple weak random collisions as a function of wavenumber, and show that the energy spectrum tends to $E(k_\perp) \propto k_\perp^{-2}$, i.e., with a spectral index of 2 [7], the same as expected from the previous scaling argument.

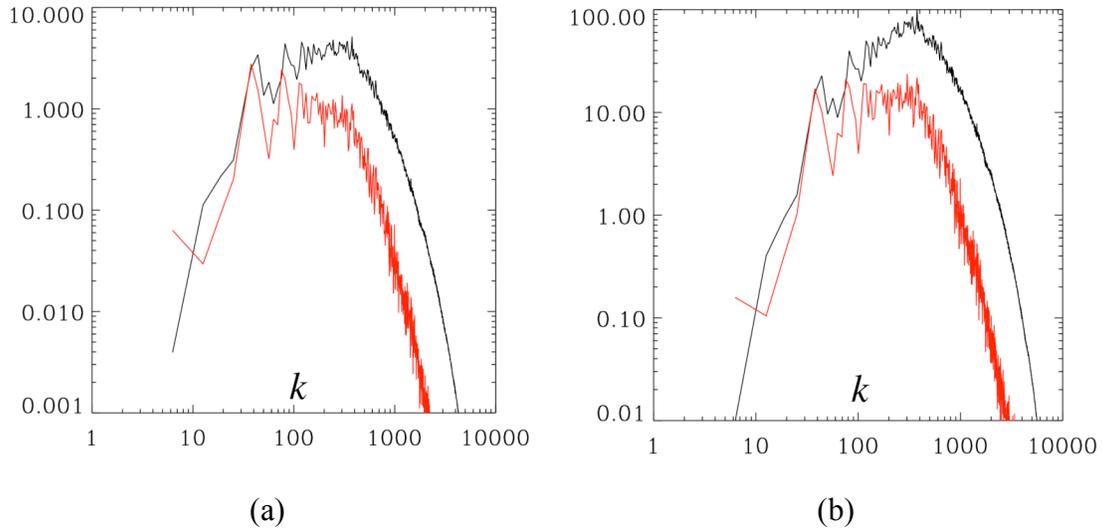

(a)  (b)

**FIGURE 1.** 1D energy spectrum $E_{1D}(k)$ (black traces) and the 2D energy spectrum $E_{2D}(k_\parallel = 0, k_\perp = k)$ along $k_\parallel = 0$ (red traces): (a) $k^{3/2} E_{1D}(k)$ and $k^{3/2} E_{2D}(k_\parallel = 0, k_\perp = k)$ (b) $k^2 E_{1D}(k)$ and $k^2 E_{2D}(k_\parallel = 0, k_\perp = k)$.

A rigorous weak turbulence theory for 3D incompressible MHD has also been developed by Galtier *et al* [8]. They derive the kinetic equations for the spectral densities of energy and helicity in the presence of a strong and uniform magnetic field $\mathbf{B}_0$. In the limit of zero cross-helicity, the exact stationary power law solution of the kinetic equation reduces to a $k_\perp^{-2}$ energy spectrum, consistent with the results discussed above. This demonstrates that the $k_\perp^{-2}$ energy spectrum, obtained from RMHD equations in [7], is not an artifact of the RMHD approximation.

In addition to the above arguments, we now present further justification from direct numerical simulations. In Fig. 1, we plot the 1D energy spectrum $E_{1D}(k)$ (black traces) and the 2D energy spectrum $E_{2D}(k_\parallel = 0, k_\perp = k)$ along $k_\parallel = 0$ (red traces) in a simulation of the 2D incompressible MHD equations with a large-scale magnetic field with its direction along the 2D plane. We see that the 1D spectrum is more consistent with $k^{-3/2}$ dependence, consistent with the isotropic IK spectrum. This is because the difficulty mentioned above of having two length-scales ($k_\perp^{-1}$ and $k_\parallel^{-1}$) in the dimensional analysis is removed after taking the integration to get the 1D energy spectrum, which only depends on $k$. Therefore, the above scaling argument can still be applied to obtain the IK spectrum with $k^{-3/2}$ dependence. Despite that, we see that the 2D energy does show a $k_\perp^{-2}$ scaling along the $k_\parallel = 0$ direction, consistent with the anisotropic weak turbulence theory presented above.

## MHD TURBULENCE WITHOUT UNIFORM MAGNETIC FIELD

In the absence of a uniform background magnetic field $\mathbf{B}_0$, it is more difficult to realize the condition for the validity of weak turbulence theory. Since the IK theory is supposed to work in both 2D and 3D, we have performed direct simulations of the 2D system in order to have higher resolutions (up to $2048^2$). We found that the energy spectra are more consistent with the IK theory, rather than Kolmogorov. In Fig. 2, we show one of such spectra, the black trace.

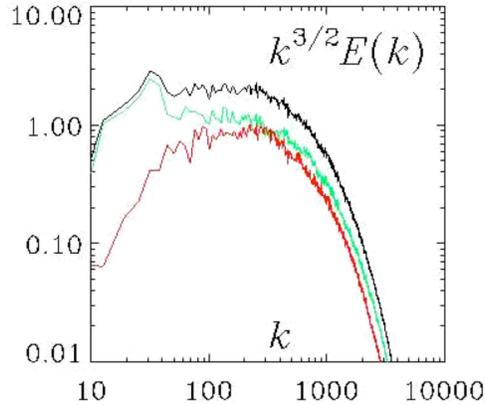

**FIGURE 2.** 1D kinetic (red), magnetic (green), and total (black) energy spectra, multiplied by a factor of $k^{3/2}$.

We have also plotted the kinetic (red) and magnetic (green) energy spectra on Fig. 2. As shown in the figure, it is generally observed that the kinetic spectrum is mostly smaller than the magnetic spectrum. The two are close to each other near the end of the inertial range and the beginning of the dissipation range. This makes the spectral index for the kinetic energy spectrum smaller than that of the total energy spectrum (i.e., 3/2) and the spectral index of the magnetic spectrum slightly above that value (since magnetic energy is larger). Interestingly, recent observations in the solar wind turbulence also show similar features [12], with the magnetic spectra closer to a Kolmogorov scaling and kinetic spectra at lower values but closer to an IK scaling.

We have also studied the local anisotropy with respect to a local magnetic field in this case and found that it is indeed locally anisotropic, as shown in Fig. 3(a), which plots the relationship (red) between parallel ($k_Z = \pi/Z$) and perpendicular ($k_R = \pi/R$) wave numbers determined by structure function method. The second-order structure function is calculated by [9, 10]

$$F_2^U(R,Z) \equiv \langle |\mathbf{U}(\mathbf{x}+\mathbf{r}) - \mathbf{U}(\mathbf{x})|^2 \rangle,$$

where $\mathbf{U}$ represents a field quantity (magnetic field or velocity field). The angle bracket indicates average over all $\mathbf{x}$ space. $Z \equiv |\mathbf{r} \cdot \hat{\mathbf{B}}_0|$ and $R \equiv |\mathbf{r} \times \hat{\mathbf{B}}_0|$ are, respectively, the parallel and perpendicular displacements relative to $\hat{\mathbf{B}}_0$, the direction of the local magnetic field. The relation between $R$ and $Z$ can be found by using the identity $F_2^U(0,Z) = F_2^U(R,0)$ which specifies the relation between the intercepts for a fixed $F_2^U$. From Fig. 3(a), we see that there is good agreement with the $k_Z \propto k_R^{2/3}$ over the inertial range, consistent with the prediction of critical balance theory [5].

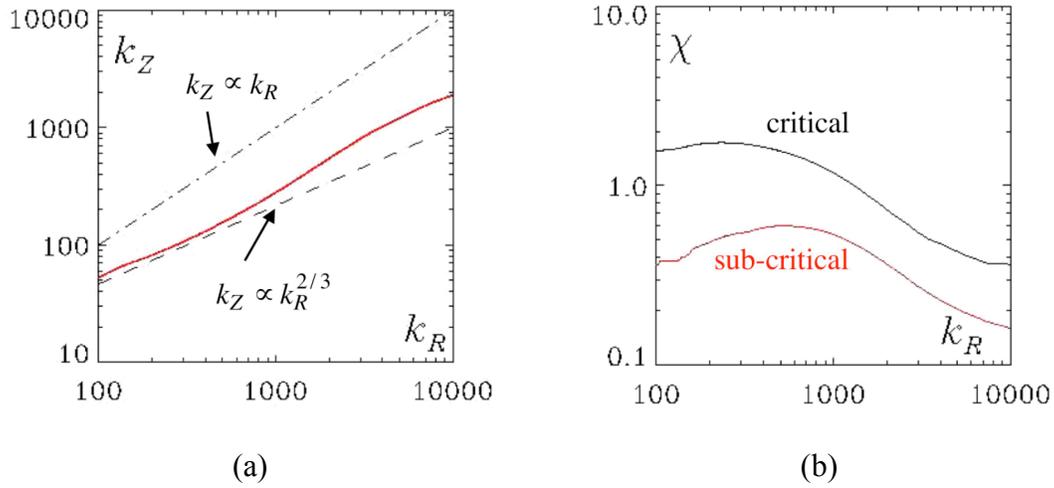

**FIGURE 3.** (a) Red trace: relationship between parallel ($k_Z$) and perpendicular ($k_R$) wave numbers determined by structure function method, showing scale-dependent anisotropy; (b) $\chi$ calculated using two different definitions showing critical ($\chi = k_\perp v_k / k_\| V_A \sim 1$, black) and sub-critical ($\chi = k_\perp v_{k\perp} / k_\| V_A < 1$, red) levels over the inertial range, in the same run as shown in Fig. 2,

However, as shown in Fig. 3(b), the critical balance condition [5] is not satisfied by about a factor of two [4], and thus the turbulence is marginally weak. The black trace is from using $\chi = k_\perp v_k / k_\| V_A$, calculated by using $\mathbf{U} = \mathbf{B}$ and $\mathbf{v}$, and is of the order of unity over the inertial range. However for the 2D case, we actually should use the definition of $\chi = k_\perp v_{k\perp} / k_\| V_A$, plotted as the red trace on Fig. 3(b) and is less than unity, by using $U = \mathbf{r} \cdot \mathbf{B}$ and $\mathbf{r} \cdot \mathbf{v}$. This is because in 2D MHD, there are no shear-Alfvén waves, and thus we do not have $v_{k\perp} \sim v_k$ as in the 3D case. In fact, if $k_\perp \gg k_\|$, we obtain $v_{k\perp} \ll v_{k\|} \sim v_k$, since $\mathbf{k} \cdot \mathbf{v} \sim k_\perp v_{k\perp} + k_\| v_{k\|} = 0$.

Although we think that the spectral index found in the simulations is closer to 3/2 than 5/3, this is numerically not definitive enough to separate the two theories, since the difference between these two values is only about 10%. Another possible way to distinguish the two theories is by looking at the scaling of the energy cascade rate $\varepsilon$ itself, since the IK rate $\varepsilon \sim k^3 E_k^2 V_A^{-1}$ is smaller than the Kolmogorov rate $\varepsilon \sim k^{5/2} E_k^{3/2}$ by the order of a small parameter $\chi = k^{1/2} E_k^{1/2} V_A^{-1}$. The energy cascade rate is straightforward for decaying turbulence, and is calculated by

$$\varepsilon = -\frac{d}{dt}\int_0^k E(k')dk' - \eta\int_0^k k'^2 E_B(k')dk' - \nu\int_0^k k'^2 E_V(k')dk' ,$$

Figure 4 shows the plots of $\varepsilon$ (black), the Kolmogorov rate $k^{5/2} E_k^{3/2}$ (blue), the IK rate $k^3 E_k^2 V_A^{-1}$ (green), and the rate of change of Fourier amplitude square (red), for cases with resolution of (a) $2048^2$ and (b) $1024^2$. At the resolution of $2048^2$, if we fit $\varepsilon$ with the Kolmogorov rate $k^{5/2} E_k^{3/2}$, we need a multiplicative constant $C_K \sim 4.6$, but if we fit it to the IK rate $k^3 E_k^2 V_A^{-1}$, the IK constant is found to be $C_{IK} \sim 1.8$. For $1024^2$, the values described above are $C_K \sim 4.0$, $C_{IK} \sim 1.8$. This suggests that if we fit $\varepsilon$ with $k^{5/2} E_k^{3/2}$, we obtain a $C_K$ increasing with the size of inertial range. However, if we fit it to $k^3 E_k^2 V_A^{-1}$, $C_{IK}$ seems to remain approximately the same. In short, it seems reasonable to conclude that the IK rate fits the data better than the Kolmogorov rate.

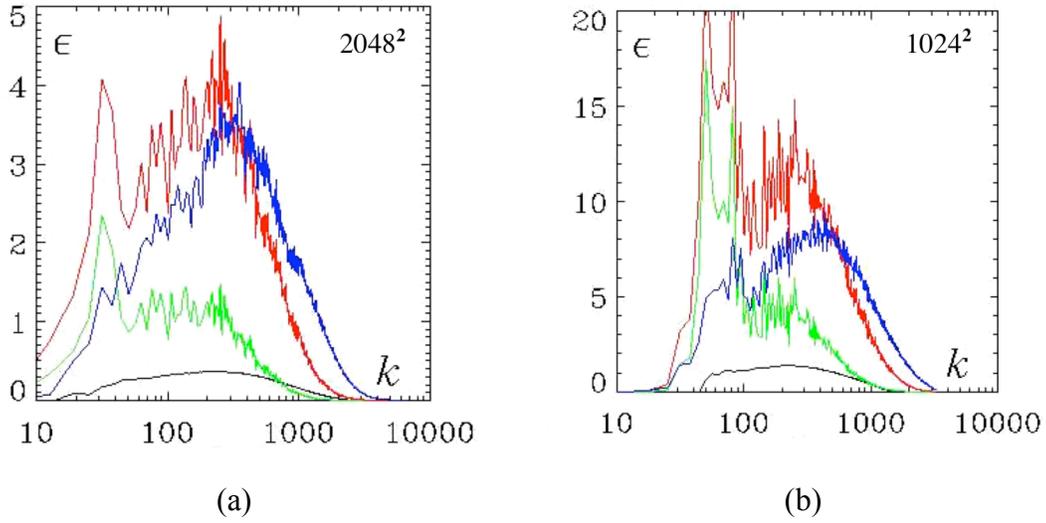

**FIGURE 4.** Numerically determined energy cascade rate $\varepsilon$ (black), the Kolmogorov rate $k^{5/2} E_k^{3/2}$ (blue), the IK rate $k^3 E_k^2 V_A^{-1}$ (green), and the rate of change of Fourier amplitude square (red), for cases with resolution of (a) $2048^2$ and (b) $1024^2$

On the other hand, the rate of change of Fourier amplitude square, defined numerically as

$$\varepsilon_A(k) \equiv \sum \left[ \left|\mathbf{v}_k(t+\Delta t) - \mathbf{v}_k(t)\right|^2 + \left|\mathbf{B}_k(t+\Delta t) - \mathbf{B}_k(t)\right|^2 \right] / 2\Delta t ,$$

is more consistent with the Kolmogorov rate. This seems to show that amplitude does change in a rate with time scale of the order of the eddy turn over time. This supposedly should give the strong turbulence limit and thus a Kolmogorov spectrum. However, such rate of change of amplitude does not seem to cascade energy directly.

Energy cascade seems to happen only after many random collisions between wave packets and thus follows the IK scaling.

Based on the numerical evidence, we conclude, with some caution, that the energy spectrum of 2D MHD turbulence is indeed consistent with a locally anisotropic IK spectrum, and that it does not attain the critical balance condition [4].

As a final remark, we note that the difference between the Kolmogorov and IK energy cascade rate can also have important implications in observations. For example, a recent study of the solar wind turbulence also shows that the observed energy cascade rate is consistent with the IK rate, but lower than the Kolmogorov rate by about an order of magnitude [13].

## CONCLUSION

In this paper, we have discussed the effects of anisotropy in MHD turbulence. For the case with a background uniform magnetic field, the weak turbulence spectrum is proportional to $k_\perp^{-2}$, as confirmed by direct simulations. We apply the concept of local anisotropy to the case without a uniform magnetic field when the turbulence is globally isotropic. For the 2D case, we confirm that there is indeed local anisotropy. However, the turbulence is found to be slightly sub-critical, which probably is due to the fact that there is no shear Alfvén wave in the 2D system. This also means that the 3D case will be more complicated, depends on the mixture of shear and pseudo Alfvén wave. The 1D energy spectral spectrum is found to be more consistent with $k^{-3/2}$, as predicted by the IK theory of weak turbulence, and with an energy cascade rate more consistent with the IK scaling.

## ACKNOWLEDGMENTS

This research is partially supported by a National Science Foundation grant AST-0434322.

## REFERENCES

1. A. N. Kolmogorov, C. R. Akad. Sci. SSSR **30**, 301 (1941).
2. P. S. Iroshnikov, Astron. Zh. **40**, 742 (1963).
3. R. H. Kraichnan, Phys. Fluids **8**, 1385 (1965).
4. C. S. Ng, A. Bhattacharjee, K. Germaschewski, and S. Galtier, Phys. Plasmas, **10**, 1954 (2003).
5. P. Goldreich and S. Sridhar, Astrophys. J. **438**, 763 (1995).
6. C. S. Ng and A. Bhattacharjee, Phys. Plasma **4**, 605 (1997).
7. A. Bhattacharjee and C. S. Ng, Astrophys. J **548**, 318 (2001).
8. S. Galtier, S. V. Nazarenko, A. C. Newell, A. Pouquet, J. Plasma Phys., **63**, 447, 2000; Astrophys. J. **564**, L49 (2002).
9. J. Cho and E. T. Vishniac, Astrophys. J. **539**, 273 (2000).
10. J. Maron and P. Goldreich, Astrophys. J. **554**, 1175 (2000).
11. C. S. Ng and A. Bhattacharjee, Astrophys. J. **465**, 845 (1996).
12. J. J. Podesta, D. A. Roberts, and M. L. Goldstein, J. Geophys. Res. **111**, A10109, doi: 10.1029/ 2006JA011834 (2006).
13. B. J. Vasquez, et al., Evaluation of the turbulent energy cascade rates from the upper inertial range in the Solar wind at 1AU, J. Geophys. Res., to appear (2007).